\documentclass[12pt]{article}
\pdfoutput=1
\usepackage{subfigure}
\usepackage{amssymb,amsmath}
\usepackage{graphicx}
\usepackage{color}
\usepackage[colorlinks=true
,urlcolor=blue
,citecolor=blue
,linkcolor=blue
,pagecolor=blue
,linktocpage=true
,pdfproducer=medialab
]{hyperref}
\usepackage[a4paper,width=15.2cm]{geometry}
\makeatletter \renewcommand{\@dotsep}{10000} \makeatother

\newcommand{\beq}{\begin{equation}}
\newcommand{\eeq}{\end{equation}}
\newcommand{\bea}{\begin{eqnarray}}
\newcommand{\eea}{\end{eqnarray}}

\begin{document}

\begin{center}

 {\Large\bf  SO(10) Yukawa Unification with $\mu < 0$
 } \vspace{1cm}

{\large   Ilia Gogoladze\footnote{E-mail: ilia@bartol.udel.edu\\
\hspace*{0.5cm} On  leave of absence from: Andronikashvili Institute
of Physics, 0177 Tbilisi, Georgia.}, Qaisar Shafi\footnote{ E-mail:
shafi@bartol.udel.edu} and Cem Salih $\ddot{\rm U}$n \footnote{
E-mail: cemsalihun@bartol.udel.edu}} \vspace{.5cm}

{\baselineskip 20pt \it
Bartol Research Institute, Department of Physics and Astronomy, \\
University of Delaware, Newark, DE 19716, USA  } \vspace{.5cm}

\vspace{1.5cm}
 {\bf Abstract}
\end{center}

We consider the low energy implications including particle spectroscopy of SO(10) inspired $t$-$b$-$\tau$ Yukawa coupling unification with $\mu < 0$, where $\mu$ is the coefficient of the bilinear Higgs mixing term of the minimal supersymmetric standard model (MSSM). We employ non-universal MSSM gaugino masses induced by SO(10) invariant dimension five operators, such that the total number of fundamental parameters is precisely the same as in Yukawa unified supersymmetric SO(10) models with universal gaugino masses and $\mu > 0$. We find that $t$-$b$-$\tau$ Yukawa unification with $\mu < 0$  is compatible with the current experimental bounds, including the WMAP bound on neutralino dark matter and the measured value of the muon anomalous magnetic moment. We present a variety of benchmark points which include relatively light squarks ($ \sim $ TeV) of the first two families and an example in which the bottom and top squarks are lighter than the gluino. This is quite distinct from Yukawa unification with $\mu > 0$  and universal gaugino masses in which the gluino is the lightest colored sparticle and the sqaurks of the first two families have masses in the multi-TeV range. 

\newpage

\renewcommand{\thefootnote}{\arabic{footnote}}
\setcounter{footnote}{0}



\section{\label{ch:introduction}Introduction}

Supersymmetric (SUSY) $SO(10)$ grand unified theory (GUT), in
contrast to its non-SUSY version, yields third family
($t$-$b$-$\tau$) Yukawa unification via the unique renormalizable
Yukawa coupling $16 \cdot 16 \cdot 10$, if the Higgs 10-plet is assumed
to contain the two Higgs doublets $H_u$ and $H_d$ of the minimal supersymmetric standard model (MSSM) \cite{big-422}. The matter 16-plet contains the 15
chiral superfields of MSSM as well as the
right handed neutrino superfield. The implications of this Yukawa unification condition at $ M_{G} \sim 2\times 10^{16}$ GeV have
been extensively explored over the years \cite{big-422,bigger-422}.
In $SO(10)$ Yukawa unification with $ \mu > 0 $ and universal gaugino masses, the gluino is the lightest colored sparticle \cite{Baer:2008jn,
Gogoladze:2009ug}, which will be tested \cite{Baer:2009ff} at the Large Hadron Collider (LHC). The
squarks and sleptons, especially those from the first two families,  turn out to have masses in the multi-TeV
range. Moreover, it is argued in \cite{Baer:2008jn,
Gogoladze:2009ug} that the lightest neutralino is not a viable cold
dark matter candidate in $SO(10)$
Yukawa unification with $ \mu > 0 $ and universal gaugino masses at $ M_{G} $

Spurred by these developments we have investigated $t$-$b$-$\tau$
Yukawa unification \cite{Gogoladze:2009ug,
Gogoladze:2009bn,Gogoladze:2010fu} in the framework of
SUSY  $SU(4)_c \times SU(2)_L \times SU(2)_R$ \cite{pati}
(4-2-2, for short). The 4-2-2 structure allows us to consider
non-universal gaugino masses while preserving Yukawa unification. An
important conclusion reached in \cite{Gogoladze:2009ug,
Gogoladze:2009bn} is that with same sign non-universal gaugino soft
terms, Yukawa unification in 4-2-2 is compatible with neutralino
dark matter, with gluino co-annihilation \cite{Gogoladze:2009ug,Baer:2009ff,
Gogoladze:2009bn, Profumo:2004wk} being a unique dark matter scenario for $ \mu > 0 $.

By considering opposite sign gauginos with
{$\mu<0,M_2<0,M_3>0$} (where $\mu$ is the coefficient of the bilinear Higgs mixing
term, $M_2$ and $M_3$ are the soft supersymmetry breaking (SSB)
gaugino mass terms corresponding respectively to $SU(2)_L$ and
$SU(3)_c$). It is shown in in \cite{Gogoladze:2010fu} that Yukawa
coupling unification consistent with the experimental constraints
can be implemented in 4-2-2. With $\mu<0$ and opposite sign gauginos,
Yukawa coupling unification is achieved for $m_0 \gtrsim 300\, {\rm
GeV}$, as opposed to $m_0 \gtrsim 8\, {\rm TeV}$ for the case of
same sign gauginos. The finite corrections to the b-quark
mass play an important role here \cite{Gogoladze:2010fu}. By considering gauginos with $M_2 <0$, $M_3>0$ and $\mu<0$,
we can obtain the correct sign for the desired contribution to
$(g-2)_\mu$ \cite{Bennett:2006fi}. This enables us to simultaneously
satisfy the requirements of  $t$-$b$-$\tau$ Yukawa unification in 4-2-2,
neutralino dark matter and $(g-2)_\mu$, as well as a variety of
other bounds.

Encouraged by the abundance of solutions and coannihilation channels
available in the case of Yukawa unified 4-2-2 with $M_2 <0$ and
$\mu<0$, it seems natural to explore Yukawa unification in SO(10) GUT (with $M_2 <0$ and $\mu<0$). It has been pointed out \cite{Martin:2009ad} that non-universal MSSM gaugino masses at $ M_{G} $ can arise from non-singlet F-terms, compatible with the underlying GUT symmetry such as SU(5) and SO(10). The SSB
gaugino masses in supergravity  \cite{Chamseddine:1982jx} can arise, say, from the following
dimension five operator:
\begin{align}
 -\frac{F^{ab}}{2 M_{\rm
Pl}} \lambda^a \lambda^b + {\rm c.c.}
\end{align}
 Here $\lambda^a$ is the two-component gaugino field, $ F^{ab} $ denotes the F-component of the field which breaks SUSY, the indices $a,b$ run over
the adjoint representation of the gauge group, and  $M_{Pl}=2.4 \times 10^{18}$ GeV is the reduced Planck mass.
  The resulting gaugino
mass matrix is $\langle F^{ab} \rangle/M_{\rm Pl}$ where the
supersymmetry breaking  parameter $\langle F^{ab} \rangle$
transforms as a singlet under the MSSM gauge group $SU(3)_{c}
\times SU(2)_L \times U(1)_Y$. The $F^{ab}$ fields belong to an
irreducible representation in the symmetric part of the direct product of the
adjoint representation of the unified group.

In SO(10), for example,
\begin{align}
({ 45} \times { 45} )_S = { 1} + { 54} + { 210} +
{ 770}
\end{align}
If  $F$  transforms as a 54 or 210 dimensional
representation of SO(10) \cite{Martin:2009ad}, one obtains the following relation
among the MSSM gaugino masses at $ M_{G} $ :
\begin{align}
M_3: M_2:M_1= 2:-3:-1 ,
\label{gaugino10}
\end{align}
where $M_1, M_2, M_3$ denote the gaugino masses of $U(1)$, $SU(2)_L$ and $SU(3)_c$
respectively. The low energy implications of this relation have recently been investigated in \cite{Okada:2011wd} without imposing Yukawa unification.

The outline for the rest of the paper is as follows.
In Section \ref{constraintsSection} we summarize the scanning procedure and the
experimental constraints that we have employed. In Section \ref{results}
we present the results from our scan and highlight
some of the predictions of an SO(10) model with $ \mu < 0 $ and the non-universal MSSM gaugino masses at $ M_{G} $ related by Eq.(\ref{gaugino10}).
 We display some benchmark points which can be tested at the LHC. Our conclusions are summarized in
Section \ref{conclusions}.

\section{Phenomenological Constraints and Scanning Procedure\label{constraintsSection}}

We employ the ISAJET~7.80 package~\cite{ISAJET}  to perform random
scans over the fundamental parameter space. In this package, the weak scale values of gauge and third generation Yukawa
couplings are evolved to $M_{\rm G}$ via the MSSM renormalization
group equations (RGEs) in the $\overline{DR}$ regularization scheme.
We do not strictly enforce the unification condition $g_3=g_1=g_2$ at $M_{\rm
G}$, since a few percent deviation from unification can be
assigned to unknown GUT-scale threshold
corrections~\cite{Hisano:1992jj}.
The deviation between $g_1=g_2$ and $g_3$ at $M_{G}$ is no
worse than $3-4\%$.
For simplicity  we do not include the Dirac neutrino Yukawa coupling
in the RGEs, whose contribution is expected to be small.

The various boundary conditions are imposed at
$M_{\rm G}$ and all the SSB
parameters, along with the gauge and Yukawa couplings, are evolved
back to the weak scale $M_{\rm Z}$.
In the evaluation of Yukawa couplings the SUSY threshold
corrections~\cite{Pierce:1996zz} are taken into account at the
common scale $M_{\rm SUSY}= \sqrt{m_{{\tilde t}_L}m_{{\tilde t}_R}}$. The entire
parameter set is iteratively run between $M_{\rm Z}$ and $M_{\rm
G}$ using the full 2-loop RGEs until a stable solution is
obtained. To better account for leading-log corrections, one-loop
step-beta functions are adopted for gauge and Yukawa couplings, and
the SSB parameters $m_i$ are extracted from RGEs at multiple scales
$m_i=m_i(m_i)$. The RGE-improved 1-loop effective potential is
minimized at $M_{\rm SUSY}$, which effectively
accounts for the leading 2-loop corrections. Full 1-loop radiative
corrections are incorporated for all sparticle masses.

The requirement of  radiative electroweak symmetry breaking (REWSB)  imposes an important theoretical
constraint on the parameter space.
In order to reconcile REWSB with Yukawa unification, the MSSM Higgs
 soft supersymmetry breaking (SSB) masses should be split in
such way  that $m^2_{H_{d}}/ m^2_{H_u}> 1.2$  at  $M_{\rm G}$ \cite{Olechowski:1994gm}.
As mentioned above, the MSSM doublets reside in the 10 dimensional
representation of SO(10) GUT  for Yukawa unification
condition to hold. In the gravity mediated
 supersymmetry breaking  scenario    \cite{Chamseddine:1982jx} the required splitting in the Higgs sector can be generated by involving additional Higgs fields \cite{Gogoladze:2011db},
  or via D-term contributions  \cite{Drees:1986vd}.
 Another important constraint
comes from limits on the cosmological abundance of stable charged
particles  \cite{Nakamura:2010zzi}. This excludes regions in the parameter space
where charged SUSY particles, such as ${\tilde \tau}_1$ or ${\tilde t}_1$, become
the lightest supersymmetric particle (LSP). We accept only those
solutions for which one of the neutralinos is the LSP and saturates
the WMAP  bound on relic dark matter abundance.

We have performed random scans for the following parameter range:
\begin{align}
0\leq  m_{0}  \leq 5\, \rm{TeV} \nonumber \\
0\leq   m_{H_u} \leq 5\, \rm{TeV} \nonumber \\
0\leq    m_{H_d} \leq 5\, \rm{TeV} \nonumber \\
0 \leq M_{1/2}  \leq 2 \, \rm{TeV} \nonumber \\
35\leq \tan\beta \leq 55 \nonumber \\
-3\leq A_{0}/m_0 \leq 3
 \label{parameterRange}
\end{align}
with  $\mu < 0$ and  $m_t = 173.1\, {\rm GeV}$  \cite{:1900yx}.
Note that our results are not
too sensitive to one or two sigma variation in the value of $m_t$  \cite{Gogoladze:2011db}.
We use $m_b(m_Z)=2.83$ GeV which is hard-coded into ISAJET.
The set of parameters presented above is usually referred to as NUHM2 \cite{Ellis:2008eu}.
This choice of parameter space  was informed by our previous works on
$t$-$b$-$\tau$ Yukawa Unification \cite{ Gogoladze:2009bn,Gogoladze:2011db}.

Employing the boundary condition from Eq.(\ref{gaugino10}) one  can define the MSSM gaugino masses at $ M_{G} $ in terms of the mass parameter $M_{1/2}$ :
\begin{align}
M_1= - M_{1/2} \nonumber \\
M_2= - 3M_{1/2} \nonumber \\
M_3=  2 M_{1/2}
 \label{gaugino11}
\end{align}

In scanning the parameter space, we employ the Metropolis-Hastings
algorithm as described in \cite{Belanger:2009ti}. The data points
collected all satisfy
the requirement of REWSB,
with the neutralino in each case being the LSP. After collecting the data, we impose
the mass bounds on all the particles \cite{Nakamura:2010zzi} and use the
IsaTools package~\cite{Baer:2002fv}
to implement the various phenomenological constraints. We successively apply the following experimental constraints on the data that
we acquire from ISAJET:
\begin{table}[h]\centering
\begin{tabular}{rlc}
$m_h~{\rm (lightest~Higgs~mass)} $&$ \geq\, 114.4~{\rm GeV}$          &  \cite{Schael:2006cr} \\
$BR(B_s \rightarrow \mu^+ \mu^-) $&$ <\, 5.8 \times 10^{-8}$        &   \cite{:2007kv}      \\
$2.85 \times 10^{-4} \leq BR(b \rightarrow s \gamma) $&$ \leq\, 4.24 \times 10^{-4} \;
 (2\sigma)$ &   \cite{Barberio:2008fa}  \\
$0.15 \leq \frac{BR(B_u\rightarrow
\tau \nu_{\tau})_{\rm MSSM}}{BR(B_u\rightarrow \tau \nu_{\tau})_{\rm SM}}$&$ \leq\, 2.41 \;
(3\sigma)$ &   \cite{Barberio:2008fa}  \\
$\Omega_{\rm CDM}h^2 $&$ =\, 0.111^{+0.028}_{-0.037} \;(5\sigma)$ &
\cite{Komatsu:2008hk} \\ $ 0 \leq \Delta(g-2)_{\mu}/2 $ & $ \leq 55.6 \times 10^{-10} $ & \cite{Bennett:2006fi}
\end{tabular}\label{table}
\end{table}


 \begin{figure}[t!]
\centering
\includegraphics[width=15.5cm]{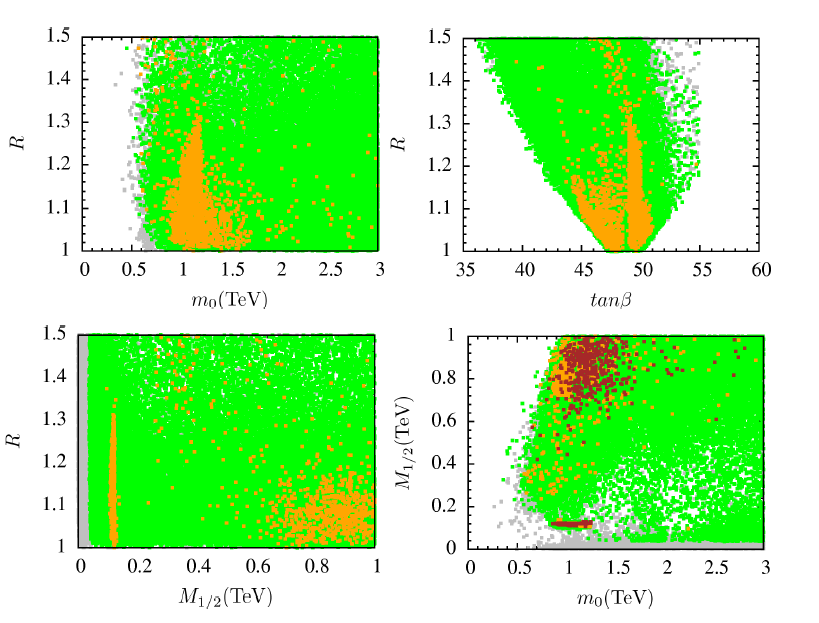}
\caption{ Plots in  $R-m_0$, $R-\tan\beta$, $R-M_{1/2}$ and $M_{1/2}-m_0$ planes.
Gray points are consistent
with REWSB and neutralino LSP.
Green points satisfy particle mass bounds and constraints from
$BR(B_s\rightarrow \mu^+ \mu^-)$, $BR(b\rightarrow s \gamma)$
and $BR(B_u\rightarrow \tau \nu_\tau)$. In addition, we require that
green points do no worse than the SM in terms of $(g-2)_\mu$.
Orange points belong to a subset of green points and satisfy
the WMAP bounds on $\tilde{\chi}^0_1$ dark matter abundance.
In the $M_{1/2}-m_0$ plane, points in brown represent a subset of yellow points and satisfy Yukawa coupling unification to within 10$\%$.
\label{fund-1}}
\end{figure}

 \begin{figure}[h!]
\centering
\includegraphics[width=15.5cm]{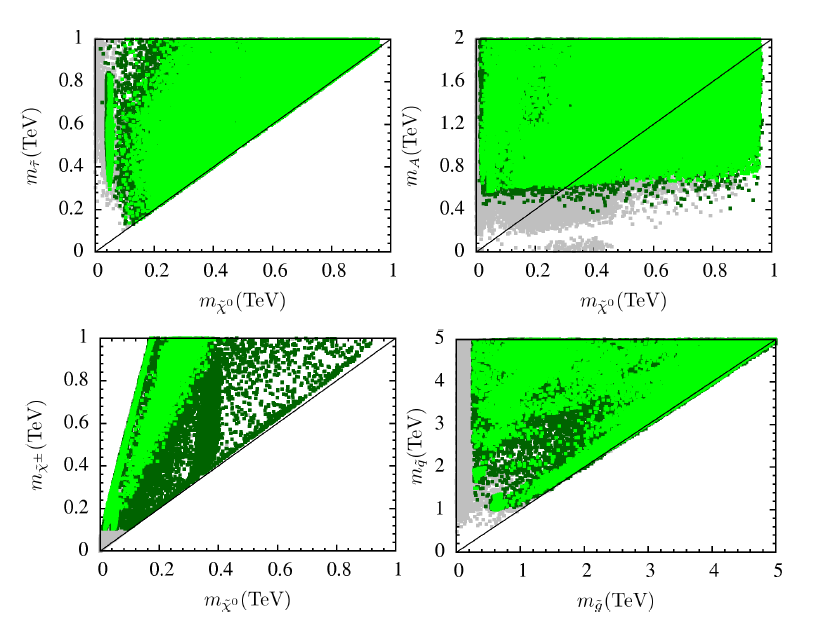}
\caption{
Plots in the $m_{\tilde
{\tau}}$ - $m_{\tilde{ \chi}_1^0}$, $m_{ {A}}$ - $m_{\tilde{ \chi}_1^0}$, $m_{\tilde{\chi}_1^{\pm}}$ - $m_{\tilde{
\chi}_1^0}$,  and $m_{\tilde{ q}}$ - $m_{\tilde{g
\chi}}$ planes. The gray points satisfy the requirements of REWSB and
$\tilde{\chi}^0_{1}$ LSP.
 Dark green points satisfy particle mass bounds and constraints from
$BR(B_s\rightarrow \mu^+ \mu^-)$, $BR(b\rightarrow s \gamma)$
and $BR(B_u\rightarrow \tau \nu_\tau)$. In addition, we require that
green points do no worse than the SM in terms of $(g-2)_\mu$. Light green points are a subset of these points which also satisfy Yukawa unification. We show in the $m_{\tilde
{\tau}}$ - $m_{\tilde{ \chi}_1^0}$ and $m_{\tilde{\chi}_1^{\pm}}$ - $m_{\tilde{
\chi}_1^0}$  planes the unit slope lines representing the respective coannihilation channels. In the
$m_A$ - $m_{\tilde{\chi}_1^0}$ plane we show the line
$m_A=2m_{\tilde{\chi}_1^0}$ that signifies the $A$ resonance
channel.
\label{spectra}}
\end{figure}


\section{Yukawa Unification and Particle Spectroscopy \label{results}}

We next present the results of the scan over the parameter space
listed in Eq.(\ref{parameterRange}). In Fig.~\ref{fund-1}  we show the
results in the  $R-m_0$, $R-\tan\beta$, $R-M_{1/2}$ and $M_{1/2}-m_0$ planes.
Gray points are consistent
with REWSB and neutralino LSP.
Green points satisfy particle mass bounds and constraints from
$BR(B_s\rightarrow \mu^+ \mu^-)$, $BR(b\rightarrow s \gamma)$
and $BR(B_u\rightarrow \tau \nu_\tau)$. In addition, we require that
green points do no worse than the SM in terms of $(g-2)_\mu$.
Orange points belong to a subset of green points and satisfy
the WMAP bounds on $\tilde{\chi}^0_1$ dark matter abundance.
In the $M_{1/2}-m_0$ plane,  points in brown represent a subset of yellow points that are consistent with Yukawa coupling unification to within 10$\%$.

In the $R$ - $m_0$ plane of Fig.~\ref{fund-1} we see that with both
$\mu<0$ and $M_2<0$, we can realize
Yukawa unification consistent with all constraints mentioned in
Section~\ref{constraintsSection} including the one from $
(g-2)_{\mu}$. This is possible because for $ \mu < 0 $, we can implement Yukawa
unification for relatively small $m_0$($\sim 500\, {\rm GeV}$),
and, in turn, $(g-2)_{\mu}$ obtains the desired SUSY contribution proportional to $\mu M_2$. This is more than an
order of magnitude reduction on the $m_0$ values required for
Yukawa unification with  $\mu>0$ and universal gaugino masses.   In the present with 10$\%$ or better  $t$-$b$-$\tau$  Yukawa  unification we obtain a relaxation also of $A_0$ values  similar to the $ SU(4)_{c}\times SU(2)_{L}\times SU(2)_{R} $ model in  \cite{Gogoladze:2010fu}, with  $-2.5<A_0/m_0<2$.  Our observation about relaxing
the possible range of $\tan\beta$ that accommodates Yukawa unified
models is explicitly shown in the $R$ - $\tan\beta$ plane.  In the $R$ - $m_{1/2}$ plane of Fig.~\ref{fund-1} we see that  employing the boundary conditions for gauginos presented in Eq.  (\ref{gaugino11}), the lightest neutralino mass can be as low as 15 GeV consistent with all constraints mentioned in
Section~\ref{constraintsSection} including the one from $
(g-2)_{\mu}$.   Note that it is impossible to realize a neutralino mass as this in the universal gaugino case due to the chargino mass constraint. A narrow orange strip for low $M_{1/2}$ values indicates the existence of $Z$ and light Higgs resonance solutions for neutralino dark matter. Actually, there are two very narrow strips, one around  45 GeV and a second around 60 GeV, even though they appear as one strip in the figure. In order to better visualize the magnitude of the sparticle masses consistent with $t$-$b$-$\tau$ Yukawa unification, we present our results in the $M_{1/2}-m_0$ plane, where the brown points correspond to Yukawa unification better than 10$\%$.

\begin{table}[]
\label{table1}
\centering
\begin{tabular}{lcccc}
\hline
\hline
                 & Point 1 & Point 2 & Point 3 & Point 4\\
\hline
$m_{0}$          & 1208    & 1027 & 1125 & 679 \\
$M_{1} $         & -677    & -111 & -122 & -216\\
$M_{2} $         & -2031 & -333 & -366 & -648\\
$M_{3} $         & 1354    & 222 & 244 & 432\\
$m_{H_{d}}$      & 1689    & 1395 & 1511 & 1263   \\
$m_{H_{u}}$      & 1260    & 1001 &  1093 & 724   \\
$\tan\beta$      & 48.1    & 49.3 & 49.6 & 48.2 \\
$A_0/m_0$        & 0.56 & -0.24   & -0.26 & 0.95 \\
$m_t$            & 173.1 & 173.1  & 173.1  & 173.1\\
\hline
$\mu$            & -938 & -246  & -276 & -263    \\

\hline
$m_h$            & 119  & 111 & 112 & 112\\
$m_H$            & 672  & 593 & 608 & 680\\
$m_A$            & 668  & 590 & 604 & 675\\
$m_{H^{\pm}}$    & 679  & 601 & 616 & 686 \\

\hline
$m_{\tilde{\chi}^0_{1,2}}$
                 & 313, 945    & 47, 211 & 52, 240 & 94, 261\\
$m_{\tilde{\chi}^0_{3,4}}$
                 &949, 1729 & 256, 333 & 286, 362 & 272, 566 \\

$m_{\tilde{\chi}^{\pm}_{1,2}}$
                 &961, 1709  & 211, 333 & 240, 363 & 263, 559 \\
$m_{\tilde{g}}$  & 2957     & 604 & 659 & 1040       \\

\hline $m_{ \tilde{u}_{L,R}}$
                 &3072, 2785  & 1142, 1108 & 1250, 1214 & 1192, 1099 \\
$m_{\tilde{t}_{1,2}}$
                 & 2197, 2602  & 691, 749 & 757, 815 & 817, 915 \\
\hline $m_{ \tilde{d}_{L,R}}$
                 & 3074, 2795  & 1145, 1131 & 1253, 1237 & 1195, 1124   \\
$m_{\tilde{b}_{1,2}}$
                 & 2227, 2585 & 630, 718 & 687, 785 & 721, 898   \\
\hline
$m_{\tilde{\nu}_{1}}$
                 & 1771       & 1034 & 1134 & 782       \\
$m_{\tilde{\nu}_{3}}$
                 &  1565      & 857  & 939  & 601     \\
\hline
$m_{ \tilde{e}_{L,R}}$
                & 1774, 1257   & 1038, 1051 & 1137, 1150 & 787, 723\\
$m_{\tilde{\tau}_{1,2}}$
                & 449, 1569  & 654, 861 & 714, 943 & 112, 607\\
\hline
$\Delta(g-2)_{\mu}$  & $0.26\times 10^{-9} $ & $0.18\times 10^{-8} $ & $0.68\times 10^{-9} $ & $ 0.19 \times 10^{-8} $       \\
$\sigma_{SI}({\rm pb})$
                & $0.56\times 10^{-9}$ & $0.12\times 10^{-7}$ & $0.86\times 10^{-8}$ & $ 0.10 \times 10^{-7} $\\

$\sigma_{SD}({\rm pb})$
                & $0.2\times 10^{-6}$ & $0.43 \times 10^{-4}$ & $0.27\times 10^{-4}$ & $ 0.38 \times 10^{-4} $\\

$\Omega_{CDM}h^{2}$
                &  0.08      & 0.104  & 0.08 & 0.12\\
\hline
$R$     &1.00 &1.02 &1.00 & 1.00\\

\hline
\hline
\end{tabular}
\caption{ Sparticle and Higgs masses (in GeV).  All of these benchmark points satisfy the various constraints
mentioned in Section~\ref{constraintsSection} and are compatible with Yukawa unification.
Point 1 depicts a solution corresponding to the A funnel region. Points 2 and 3
display the light Higgs and Z-resonance solutions, while point 4 represents the stau coannihilation solution.
\label{table1}}
\end{table}


In Fig.~\ref{spectra} we show the relic density channels consistent
with Yukawa unification in the $m_{\tilde
{\tau}}$ - $m_{\tilde{ \chi}_1^0}$, $m_{ {A}}$ - $m_{\tilde{ \chi}_1^0}$, $m_{\tilde{\chi}_1^{\pm}}$ - $m_{\tilde{
\chi}_1^0}$,  and $m_{\tilde{ q}}$ - $m_{\tilde{g}}$  planes. Gray points shown in this figure
satisfy the requirements of REWSB and $\tilde{\chi}^0_{1}$ LSP.
 Dark green points satisfy the particle mass bounds and constraints from
$BR(B_s\rightarrow \mu^+ \mu^-)$, $BR(b\rightarrow s \gamma)$
and $BR(B_u\rightarrow \tau \nu_\tau)$. the green points do no worse than the SM in terms of $(g-2)_\mu$.
The light green points represent a subset of the dark green points, and correspond to 10$ \% $ or better  $t$-$b$-$\tau$ Yukawa unification. This choice of color coding is
influenced from displaying the sparticle spectrum with and without
$t$-$b$-$\tau$ Yukawa unification, while still focussing on all the other
experimental constraints. The idea is to show the myriad of
solutions that implement Yukawa unification and are consistent with
all known experimental bounds except that on relic dark matter
density from WMAP. The appearance of a variety of Yukawa unified solutions
with a very rich sparticle spectrum is a characteristic feature of $\mu<0$  \cite{Gogoladze:2010fu}.

We can see in Fig.~\ref{spectra} that a variety of coannihilation
and resonance scenarios are compatible with Yukawa unification
and neutralino dark matter. Included in the $m_A$ - $m_{\tilde{
\chi}_1^0}$ plane is the line $m_A$ = $2 m_{\tilde{ \chi}_1^0}$
which shows that the $A$-funnel region is compatible
with Yukawa unification.
In the $m_{\tilde
{\tau}}$ - $m_{\tilde{ \chi}_1^0}$  plane in
Fig.~\ref{spectra}, we draw the unit slope line which indicates the
presence of  stau  coannihilation scenarios.
From the $m_{\tilde{\chi}_1^{\pm}}$ - $m_{\tilde{
\chi}_1^0}$ plane, it is easy to recognize the light Higgs ($h$) and
$Z$ resonance channels.  We expect  that other coannihilation channels like the
stop coannihilation scenario are also consistent with Yukawa
unification, although we have not found them, perhaps due to lack of statistics.

Let us remark on the low mass neutralino solutions
that we have found in our model (with $ \mu < 0 $).
Because of the $M_{\rm G}$ scale  gaugino mass relations in Eq.(\ref{gaugino11}),  it is
possible in principle to have small $M_1$ values, thus giving rise to a light neutralino.  The
neutralino mass nonetheless is bounded from below because of the
relic density bounds on dark matter. The SO(10) model with non-universal gaugino masses, as in this paper, has all the ingredients to bring down the neutralino
mass to the lowest possible value consistent with the various constraints. The solution with the neutralino (mass $ \sim 43 $ GeV) is
consistent with Yukawa unification and corresponds the Z-resonance dark matter scenario.

Finally, in Table~\ref{table1} we present some benchmark points for the SO(10) $t$-$b$-$\tau$ Yukawa unified
model with $\mu<0$ and non-universal gaugino masses. All of these points contain
WMAP compatible with neutralino dark matter and satisfy the constraints
mentioned in Section~\ref{constraintsSection}. Point 1 depicts a solution with essentially perfect Yukawa unification corresponding to the A funnel region. Points 2 and 3
correspond to the light Higgs and Z-resonance solutions, while point 4 represents the stau coannihilation solution.
It is interesting to note that for the light Higgs and Z-resonance solutions, there is an upper bound on the gluino mass. Employing the boundary condition in Eq.(\ref{gaugino11}) this turns out to be $m_{g}\approx 700$ GeV. Hence, the light Higgs and Z-resonance solutions are not compatible with this model if the gluinos are founded to be heavier than $ \approx 700 $ GeV.


\section{Conclusion \label{conclusions}}

We have shown that SO(10) $t$-$b$-$\tau$ Yukawa unification with $\mu < 0$  and non-universal gaugino masses is nicely consistent with all available experimental data. We have considered a variety of WMAP compatible neutralino dark matter scenarios, including some examples in which the LSP neutralino can be rather light, about half the Z-boson mass (Z-resonance solution) or the SM-like Higgs mass (light Higgs resonance solution). Neutralino dark matter solutions corresponding to the A-funnel region and stau-coannihilation are also shown to exist. With $\mu M_2 > 0$, the SUSY contributions to the muon anomalous magnetic moment can help provide better agreement than the SM with the experimental data. Finally, in comparison to SO(10) with $\mu > 0$  and universal gaugino masses, there exist some important differences, even though the number of fundamental parameters in the two cases are the same. The lack of WMAP compatible neutralino dark matter in the $\mu > 0$  case is one of them. Also, with $\mu < 0$, we find examples in which the first two squark families are relatively light ($ \sim $ TeV), and the third family $b$ and $t$ squarks can be lighter than the gluino (which happens to be the lightest colored sparticle in SUSY SO(10) with $\mu > 0$ ).

\textbf{Note Added:} As we were finishing this work, Stuart Raby pointed out that M. Badziak, M Olechowski and S. Pokorski are also investigating SO(10) Yukawa unification with $ \mu < 0 $

\section*{Acknowledgments}
We thank Stuart Raby and Shabbar Raza  for valuable discussions.
I.G.  wishes to thank the  Center for Theoretical Underground Physics and Related Areas (CETUP*)  where some part  of this project was done.
This work is supported in part by the DOE Grant No. DE-FG02-91ER40626 (I.G.,
C.U., and Q.S.).



\begin{thebibliography}{99}

\bibitem{big-422}
B. Ananthanarayan, G. Lazarides and Q. Shafi, Phys. Rev. D {\bf 44},
1613 (1991) and Phys. Lett. B {\bf 300}, 24 (1993)5; Q.~Shafi and
B.~Ananthanarayan, Trieste HEP Cosmol.1991:233-244.

\bibitem{bigger-422}
L.~J.~Hall, R.~Rattazzi and U.~Sarid, Phys.\ Rev.\  D {\bf 50}, 7048 (1994);
M. Olechowski and S. Pokorski, Phys.\ Lett.\  B {\bf 214}, 393 (1988);
T. Banks, Nucl.\ Phys.\ B {\bf 303}, 172 (1988);
V. Barger, M. Berger and P. Ohmann, Phys. Rev. D {\bf 49}, (1994)
4908; M. Carena, M. Olechowski, S. Pokorski and C. Wagner,  Nucl.\
Phys.\  B {\bf 426}, 269 (1994); B. Ananthanarayan, Q. Shafi and X.
Wang, Phys. Rev. D {\bf 50}, 5980 (1994); G. Anderson et al. Phys.
Rev. D {\bf 47}, (1993) 3702 and Phys. Rev. D {\bf 49},  3660
(1994); R. Rattazzi and U. Sarid, Phys. Rev. D {\bf 53}, 1553
(1996); T. Blazek, M. Carena, S. Raby and C. Wagner, Phys. Rev. D
{\bf 56}, 6919 (1997); T. Blazek, S. Raby and K. Tobe, Phys. Rev. D
{\bf 62}, 055001 (2000); H. Baer, M. Diaz, J. Ferrandis and X. Tata,
Phys. Rev. D {\bf 61}, 111701 (2000); H. Baer, M. Brhlik, M. Diaz,
J. Ferrandis, P. Mercadante, P. Quintana and X. Tata, Phys. Rev. D
{\bf 63}, 015007(2001);  C.~Balazs and R.~Dermisek, JHEP {\bf 0306}, 024 (2003);
C. Pallis, Nucl. Phys. B {\bf 678},  398 (2004); U. Chattopadhyay, A. Corsetti and P.
Nath, Phys. Rev. D {\bf 66} 035003, (2002); T.~Blazek, R.~Dermisek
and S.~Raby, Phys.\ Rev.\ Lett.\  {\bf 88}, 111804 (2002) and Phys.\
Rev.\  D {\bf 65}, 115004 (2002); M. Gomez, T. Ibrahim, P. Nath and
S. Skadhauge, Phys. Rev. D {\bf 72}, 095008 (2005); K. Tobe and J.
D. Wells, Nucl. Phys. B {\bf 663}, 123 (2003);
I.~Gogoladze, Y.~Mimura, S.~Nandi,
  Phys.\ Lett.\  {\bf B562}, 307 (2003);
W.~Altmannshofer,
D.~Guadagnoli, S.~Raby and D.~M.~Straub, Phys.\ Lett.\  B {\bf 668},
385 (2008);
S.~Antusch and M.~Spinrath,
 Phys.\ Rev.\  D {\bf 78}, 075020 (2008);
 H.~Baer, S.~Kraml and S.~Sekmen, JHEP {\bf 0909}, 005 (2009);
S.~Antusch and M.~Spinrath,
Phys.\ Rev.\  D {\bf 79}, 095004 (2009);
D.~Guadagnoli, S.~Raby and D.~M.~Straub, JHEP {\bf 0910}, 059 (2009);
K.~Choi, D.~Guadagnoli, S.~H.~Im and C.~B.~Park,
  JHEP {\bf 1010}, 025 (2010);
 S.~Dar, I.~Gogoladze, Q.~Shafi and C.~S.~Un,
  arXiv:1105.5122 [hep-ph];
  N.~Karagiannakis, G.~Lazarides and C.~Pallis,
  arXiv:1107.0667 [hep-ph].



\bibitem{Baer:2008jn}
 H.~Baer, S.~Kraml, S.~Sekmen and H.~Summy,
  JHEP {\bf 0803}, 056 (2008);
 H.~Baer, M.~Haider, S.~Kraml, S.~Sekmen and H.~Summy,
  JCAP {\bf 0902}, 002 (2009).


\bibitem{Gogoladze:2009ug}
  I.~Gogoladze, R.~Khalid and Q.~Shafi,
  Phys.\ Rev.\  D {\bf 79}, 115004 (2009).


\bibitem{Baer:2009ff}
 H.~Baer, S.~Kraml, A.~Lessa and S.~Sekmen,
 JHEP {\bf 1002}, 055 (2010);
  D.~Feldman, Z.~Liu and P.~Nath,
  Phys.\ Rev.\  D {\bf 80}, 015007 (2009);
  M.~A.~Ajaib, T.~Li, Q.~Shafi and K.~Wang,
  JHEP {\bf 1101}, 028 (2011).

\bibitem{Gogoladze:2009bn}
  I.~Gogoladze, R.~Khalid and Q.~Shafi,
  Phys.\ Rev.\  D {\bf 80}, 095016 (2009);


\bibitem{Gogoladze:2010fu}
  I.~Gogoladze, R.~Khalid, S.~Raza and Q.~Shafi,
  JHEP {\bf 1012}, 055 (2010);
  arXiv:1008.2765 [hep-ph].

\bibitem{pati}
J.~C.~Pati and A.~Salam,
  Phys.\ Rev.\  D {\bf 10}, 275 (1974).

\bibitem{Profumo:2004wk}
  S.~Profumo and C.~E.~Yaguna,
  Phys.\ Rev.\  D {\bf 69}, 115009 (2004);
  D.~Feldman, Z.~Liu and P.~Nath,
  Phys.\ Rev.\  D {\bf 80}, 015007 (2009).





\bibitem{Bennett:2006fi}
  G.~W.~Bennett {\it et al.}  [Muon G-2 Collaboration],
  Phys.\ Rev.\  D {\bf 73}, 072003 (2006).

\bibitem{Martin:2009ad}
 See, for instance,  S.~P.~Martin,
  Phys.\ Rev.\  {\bf D79}, 095019 (2009);
    U.~Chattopadhyay, D.~Das and D.~P.~Roy,
  Phys.\ Rev.\  D {\bf 79}, 095013 (2009);
   B.~Ananthanarayan, P.~N.~Pandita,
  Int.\ J.\ Mod.\ Phys.\  {\bf A22}, 3229-3259 (2007);
  S.~Bhattacharya, A.~Datta and B.~Mukhopadhyaya,
  JHEP {\bf 0710}, 080 (2007);
  A.~Corsetti and P.~Nath,
  Phys.\ Rev.\  D {\bf 64}, 125010 (2001)
   and references therein.




\bibitem{Chamseddine:1982jx}
 A.~Chamseddine, R.~Arnowitt and P.~Nath, Phys.\ Rev.\ Lett.\ {\bf 49} (1982) 970;
R.~Barbieri, S.~Ferrara and C.~Savoy, Phys.\ Lett.\ {\bf B119}
(1982) 343; N.~Ohta, Prog.\ Theor.\ Phys.\ {\bf 70} (1983) 542;
L.~J.~Hall, J.~D.~Lykken and S.~Weinberg, Phys.\ Rev.\ {\bf D27}
(1983) 2359; for a review see  S.~Weinberg, {\it The Quantum Theory
of Fields: Volume 3, Supersymmetry,
 Cambridge University Press (2000) 442p}.

\bibitem{Okada:2011wd}
  N.~Okada, S.~Raza and Q.~Shafi,
  arXiv:1107.0941 [hep-ph].


\bibitem{ISAJET}
H.~Baer, F.~E.~Paige, S.~D.~Protopopescu and X.~Tata,
arXiv:hep-ph/0001086.

\bibitem{Hisano:1992jj}
J.~Hisano, H.~Murayama, and T.~Yanagida,
  { Nucl. Phys.} {\bf B402} (1993) 46.
Y.~Yamada,
{ Z. Phys.} {\bf C60} (1993) 83;
 J.~L.~Chkareuli and I.~G.~Gogoladze,
  Phys.\ Rev.\  D {\bf 58}, 055011 (1998).





\bibitem{Pierce:1996zz}
D.~M. Pierce, J.~A. Bagger, K.~T. Matchev, and R.-j. Zhang,
  { Nucl. Phys.} {\bf B491} (1997) 3.


\bibitem{Olechowski:1994gm}
  M.~Olechowski and S.~Pokorski,
  Phys.\ Lett.\  B {\bf 344}, 201 (1995);
  D.~Matalliotakis and H.~P.~Nilles,
  Nucl.\ Phys.\  B {\bf 435}, 115 (1995);
  H.~Murayama, M.~Olechowski and S.~Pokorski,
  Phys.\ Lett.\  B {\bf 371}, 57 (1996).



\bibitem{Gogoladze:2011db}
  I.~Gogoladze, R.~Khalid, S.~Raza and Q.~Shafi,
  JHEP {\bf 1106} (2011) 117.

\bibitem{Drees:1986vd}
  M.~Drees,
  Phys.\ Lett.\  B {\bf 181}, 279 (1986);
  C.~F.~Kolda and S.~P.~Martin,
  Phys.\ Rev.\  D {\bf 53}, 3871 (1996)
  [arXiv:hep-ph/9503445].




\bibitem{Nakamura:2010zzi}
  K. Nakamura {\it et al.} [ Particle Data Group Collaboration ],
  J.\ Phys.\ G {\bf G37}, 075021 (2010).

\bibitem{:1900yx}
 [Tevatron Electroweak Working Group and CDF Collaboration and D0 Collab],
  arXiv:0903.2503 [hep-ex].




\bibitem{Ellis:2008eu} See for instance
  J.~R.~Ellis, K.~A.~Olive, P.~Sandick,
  Phys.\ Rev.\  {\bf D78}, 075012 (2008) and references therein.




\bibitem{Belanger:2009ti}
  G.~Belanger, F.~Boudjema, A.~Pukhov and R.~K.~Singh,
  JHEP {\bf 0911}, 026 (2009);
H.~Baer, S.~Kraml, S.~Sekmen and H.~Summy,
  JHEP {\bf 0803}, 056 (2008).

\bibitem{Baer:2002fv}
H.~Baer, C.~Balazs, and A.~Belyaev,
   { JHEP} {\bf 03} (2002) 042;
 H.~Baer, C.~Balazs, J.~Ferrandis, and X.~Tata
  { Phys. Rev.} {\bf D64} (2001)  035004.

\bibitem{Schael:2006cr}
  S.~Schael {\it et al.}  
  Eur.\ Phys.\ J.\  C {\bf 47}, 547 (2006).


\bibitem{:2007kv}
  T.~Aaltonen {\it et al.}  [CDF Collaboration],
  Phys.\ Rev.\ Lett.\  {\bf 100}, 101802 (2008).


\bibitem{Barberio:2008fa}
  E.~Barberio {\it et al.}  [Heavy Flavor Averaging Group],
  arXiv:0808.1297 [hep-ex].




\bibitem{Komatsu:2008hk}
  E.~Komatsu {\it et al.}  [WMAP Collaboration],
  Astrophys.\ J.\ Suppl.\  {\bf 180}, 330 (2009).







\end{thebibliography}
\end{document}